\documentclass[twocolumn,english]{revtex4}
\usepackage[T1]{fontenc}
\usepackage[latin9]{inputenc}
\usepackage{textcomp}
\usepackage{amstext}
\usepackage{graphicx}

\makeatletter

\newcommand{\lyxmathsym}[1]{\ifmmode\begingroup\def\b@ld{bold}
  \text{\ifx\math@version\b@ld\bfseries\fi#1}\endgroup\else#1\fi}

\@ifundefined{textcolor}{}
{%
 \definecolor{BLACK}{gray}{0}
 \definecolor{WHITE}{gray}{1}
 \definecolor{RED}{rgb}{1,0,0}
 \definecolor{GREEN}{rgb}{0,1,0}
 \definecolor{BLUE}{rgb}{0,0,1}
 \definecolor{CYAN}{cmyk}{1,0,0,0}
 \definecolor{MAGENTA}{cmyk}{0,1,0,0}
 \definecolor{YELLOW}{cmyk}{0,0,1,0}
 }

\makeatother

\usepackage{babel}
\begin{document}

\title{Magnetic force microscopy study of the switching field distribution
of low density arrays of single domain magnetic nanowires}

\author{M. R. Tabasum$^{1}$, F. Zighem$^{1,2}$, J. De La Torre Medina$^{3}$,
A. Encinas$^{4}$, L. Piraux$^{1}$ and B. Nysten$^{1}$}

\affiliation{$^{1}$Institute of Condensed Matter and Nanosciences (IMCN), Université
catholique de Louvain, \textit{Belgium}}

\affiliation{$^{2}$Laboratoire des Sciences des Procédés et des Matériaux, CNRS
- Université Paris 13, \textit{France}}

\affiliation{$^{3}$ Facultad de Ciencias Físico-Matemáticas, Universidad Michoacana
de San Nicolás de Hidalgo, Morelia, \textit{Mexico}}

\affiliation{$^{4}$ Instituto de Física, Universidad Autónoma de San Luis Potosí,
\textit{Mexico}}
\begin{abstract}
In the present work, we report on the in situ magnetic force microscopy
(MFM) study of the magnetization reversal in two-dimensional arrays
of ferromagnetic Ni$_{80}$Fe$_{20}$ and Co$_{55}$Fe$_{45}$ nanowires
(NW) with different diameters (40, 50, 70 and 100 nm) deposited inside
low porosity ($P<1\%$) nanoporous polycarbonate membranes. In such
arrays, the nanowires are sufficiently isolated from each other so
that long range dipolar interactions can be neglected. The MFM experiments
performed for different magnetization states at the same spot of the
samples are analysed to determine the switching field distribution
(SFD). The magnetization curves obtained from the MFM images are relatively
square shaped. The SFD widths are narrower compared to those obtained
for high density arrays. The weak broadening of the curves may be
ascribed to the NW intrinsic SFD. The influence of diameter and composition
of the ferromagnetic NW is also investigated.
\end{abstract}

\keywords{Polarized SANS, form factors, micromagnetic simulations }

\maketitle

\section{Introduction }

Ordered arrays of isolated magnetic nanostructures are of considerable
interest for replacing continuous film magnetic recording media for
ultrahigh density beyond 1 Tbit/in$^{2}$. In such patterned media,
each artificially fabricated magnetic nanostructure is capable of
storing an individual bit. In order to make patterned media technologically
viable a strict control of the switching field distribution (SFD)
is essential {[}1, 2{]}. In recent years, there have been a considerable
number of studies regarding the formation of arrays of single domain
magnetic nanowires (NW) embedded in nanoporous templates by electro
deposition {[}3-8{]}. The large aspect ratio of these nanowires ensures
perpendicular nanostructured media due to shape anisotropy. Although
the SFD can be obtained from a DC demagnetization remanence curve
averaged over a large nanowire array, magnetic force microscopy (MFM)
has been found to be an interesting approach to probe locally the
magnetization reversal process and to obtain local hysteresis loops
{[}9-12{]}. In some conditions, the MFM may also be used to locally
manipulate the magnetization in arrays of ferromagnetic NW {[}13{]}.
Also, interesting two dimensional labyrinth pattern of magnetic domain
structure in the demagnetized state was demonstrated using MFM measurements
in hexagonal lattices formed in anodized aluminum oxide (AAO) templates
{[}9-12{]}.

The SFD has two components that should be addressed separately {[}12,
14, 15{]}: an intrinsic part and a dipolar contribution arising from
the magnetostatic interaction of a nanowire with its neighbours. In
the latter case, the reversal field of a nanowire depends on the magnetic
state of its neighbours. Besides, there are many causes for intrinsic
SFD that may arise from local variations of the nanowire magnetic
properties and disparity in nanowire sizes and edge effects. Such
variations manifest themselves as a distribution of the switching
fields associated to the magnetization reversal process of nanowires.
From previous studies performed on dense arrays of single domain nanowires
embedded in AAO templates, it appears that the SFD is quite large,
up to 5-8 kOe {[}3, 12, 16{]}, resulting from the large dipolar magnetic
interaction that dominates the behaviour of hysteresis loops. Discrimination
between intrinsic and dipolar contributions is not straightforward
and requires dipole-dipole interaction models to estimate the intrinsic
SFD {[}12, 17{]}. 

In this work, we present a fundamental MFM study to evaluate the intrinsic
SFD in arrays of bistable Ni$_{80}$Fe$_{20}$ and Co$_{55}$Fe$_{45}$
NW deposited in low porosity polycarbonate (PC) membranes with controlled
diameters in the range 40-100 nm. All NW can be considered as infinite
long cylinders. It is shown that the intrinsic SFD, extracted directly
from the MFM measurements without involving the complicated calculations,
is considerably reduced. In addition, the non-uniformity in wire diameter
manifest itself as an increase of the intrinsic SFD as the diameter
is reduced.

\section{Experimental procedures}

\subsection{Fabrication of arrays of magnetic nanowires }

The arrays of Ni$_{80}$Fe$_{20}$ and Co$_{55}$Fe$_{45}$ NW have
been fabricated by electrodeposition in track-etched PC membranes
of porosity (P) less than $1
$. For this study, membranes with pore diameters ($D$) of 40, 50,
70 and 100 nm have been used. The experimental procedure for obtaining
the nanoporous PC membranes used in this study is detailed elsewhere
{[}18{]}. From scanning electron microscopy (SEM) measurements (not
shown here), the diameter distribution of the membrane pores has a
standard deviation $\sigma_{D}=\pm5$ nm. 

Prior to the electrodeposition, a gold layer was evaporated on one
side of the membrane in order to cover the pores and use it as a cathode.
For proper adhesion of the gold layer, a thin layer of Cr ($\sim$10
nm) was first deposited on the PC membranes. Ni$_{80}$Fe$_{20}$
NW were grown at a constant potential of $-1.05$ V from an electrolyte
containing NiSO$_{4}$.6H$_{2}$O (131 g/L), FeSO$_{4}$.6H$_{2}$O
(5.56 g/L) and H$_{3}$BO$_{3}$ (24.7 g/L). The Co$_{55}$Fe$_{45}$
NW were deposited at a constant potential of $-0.90$ V from a CoSO$_{4}$.7H$_{2}$O
(80 g/L), FeSO$_{4}$.6H$_{2}$O (40 g/L), H$_{3}$BO$_{3}$ (30 g/L)
electrolyte at room temperature. The high aspect ratio NW presented
in this study have a mean length ($L$) of around 4 $\lyxmathsym{\textmu}$m
with a standard deviation of $\pm0.1$ \textmu{}m measured using SEM.
Furthermore, previous studies showed that Ni$_{80}$Fe$_{20}$ and
Co$_{55}$Fe$_{45}$ NW synthesized in such conditions present a polycrystalline
cubic structure with no preferred texture along the NW axis, so that
the magnetocrystalline anisotropy can be neglected {[}11, 19, 20,
21{]}. The effective packing density $P$, defined as the total area
of the top end of the NW $S_{NW}$ divided by the total area of the
surface of membrane under consideration $S_{tot}$: $P=\frac{S_{NW}}{S_{tot}}$
and the mean inter-wire distances $I_{D}$, for the first four neighbouring
wires, have been measured from MFM images.

\subsection{MFM experiments}

A smooth surface, where all the nanowire tips are close to the surface
at one side, has been obtained by removing the Au and Cr layers using
a chemical etching procedure. MFM experiments were then performed
under ambient conditions using an Agilent 5500 microscope (Agilent
Technologies) equipped with 100 $\lyxmathsym{\textmu}$m closed-loop
scanner. The MFM probes, Bruker MESP-HM with a force constant around
2.8 N.m$^{-1}$ and a resonance frequency of about 75 kHz were used
for this study. The analyses were realised in amplitude-modulation
(AM-AFM) using a double pass procedure. First, the topography of one
line was recorded in standard intermittent-contact mode. Then, the
probe was lifted up a few tens of nanometres and the same line was
scanned at constant probe-surface distance; the phase signal proportional
to the magnetic interaction was simultaneously recorded.

The setup of the instrument was modified with a custom-built electromagnet
for performing \textit{in situ} MFM experiments allowing the analysis
of the same spots of the samples after applying various magnetic fields
parallel to the NW direction. Before inserting the sample inside the
setup, the NW array magnetization was saturated along their axis ($+O$z)
under a magnetic field of $H=+2$ kOe while the MFM probe tip was
saturated in the opposite direction ($-Oz$). For the MFM images,
phase-lag has been measured which is inversely proportional to the
force gradient as shown in the equation below.
\[
\Delta\phi\propto\frac{-Q}{k}\frac{\partial F}{\partial z}
\]

$Q$ is the cantilever quality factor and $k$ its stiffness. Therefore,
attractive forces with a positive gradient lead to a negative phase-lag
(dark contrast) and repulsive forces with a negative gradient lead
to a positive phase-lag (bright contrast). The different MFM-images
presented in this work were obtained at zero fields, in the remnant
states, after applying increasing reverse magnetic fields along the
NW\textquoteright{}s axis. In addition to the MFM experiments, magnetization
curves were obtained using an alternating gradient field magnetometer
(AGFM). The FEMM package {[}21{]} has been used in order to estimate
the dipolar field created by an isolated and uniformly magnetized
NW. 

\begin{figure}
\includegraphics[bb=30bp 220bp 440bp 570bp,clip,width=8.5cm]{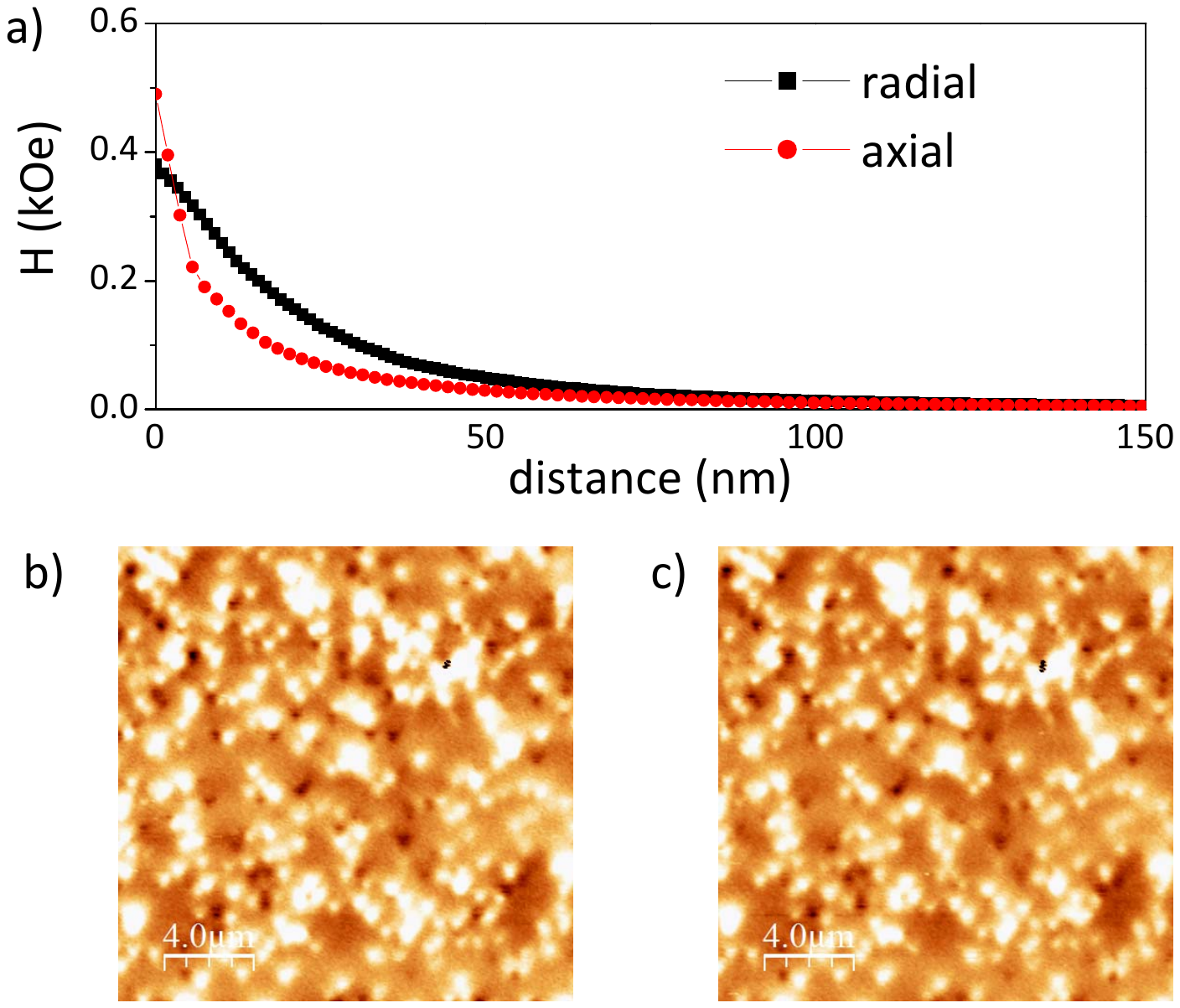}
\caption{a) Simulation of the radial and axial components of the stray field
from an isolated nanowire ends using FEMM {[}21{]} and considering
a 4 $\lyxmathsym{\textmu}$m-long NW of diameter $D=100$ nm uniformly
magnetized. The magnetic parameters used in the simulation correspond
to the Ni$_{80}$Fe$_{20}$ composition ($M_{S}=850$ emu.cm$^{-3}$).
The average inter-wire distance for the 100 nm diameter NW array is
600 nm and 950 nm for Co$_{55}$Fe$_{45}$ and Ni$_{80}$Fe$_{20}$
respectively. MFM image of the Co$_{55}$Fe$_{45}$ nanowire array
($D=50$ nm) in (b) a positive 0.6 kOe field and (c) lowering the
field to zero to verify the bistable behaviour of the nanowire array.}

\label{Fig1:images} 
\end{figure}

\section{Results and discussion}

The $I_{D}$ between nanowires was found to be larger than 600 nm
(measured between the first four neighboring wires). Figure 1a presents
the dipolar field created by a uniformly magnetized and isolated NW
($L=4$ $\lyxmathsym{\textmu}$m, $D=100$ nm) using the Ni$_{80}$Fe$_{20}$
saturation magnetization value: $M_{S}=850$ emu.cm$^{-3}$. It illustrates
that for $I_{D}>$ 600 nm the dipolar interactions between NW can
be neglected. As a consequence, the magnetic state in the remnant
state and under an applied magnetic field should be the same. This
is demonstrated by the MFM images of a Co$_{55}$Fe$_{45}$ nanowire
array ($D=5$0 nm) in a positive 600 Oe field (Figure 1b) and as the
field was lowered to zero (Figure 1c). It can be observed that both
images are similar demonstrating that the low packing density of the
NW array indeed leads to weak dipolar interactions between NWs, and
that measurements done at a field $H_{0}$ and at zero applied field
after applying a field $H_{0}$ are same.

Figures 2a-b present typical MFM images obtained for the array of
Co$_{55}$Fe$_{45}$ NW ($D=70$ nm) and Ni$_{80}$Fe$_{20}$ NW ($D=70$
nm). The first image (top left) corresponds to a situation where all
the NW are uniformly magnetized in a field $H_{0}=+2000$ Oe along
+Oz while the tip is magnetized along $-Oz$. Then, a series of magnetic
fields opposite to the sample initial saturation field were applied.
The successive switching of the NW started after the application of
around 360 Oe and 310 Oe until the application of around 1200 Oe and
585 Oe for Co$_{55}$Fe$_{45}$ and Ni$_{80}$Fe$_{20}$ NW, respectively.
The spot sizes (black or white) are larger than the nominal diameters
of the wires because MFM measures the dispersive stray field emanating
from the wires and not the exact wire dimensions.

\begin{figure}
\includegraphics[bb=75bp 0bp 460bp 570bp,clip,width=8.5cm]{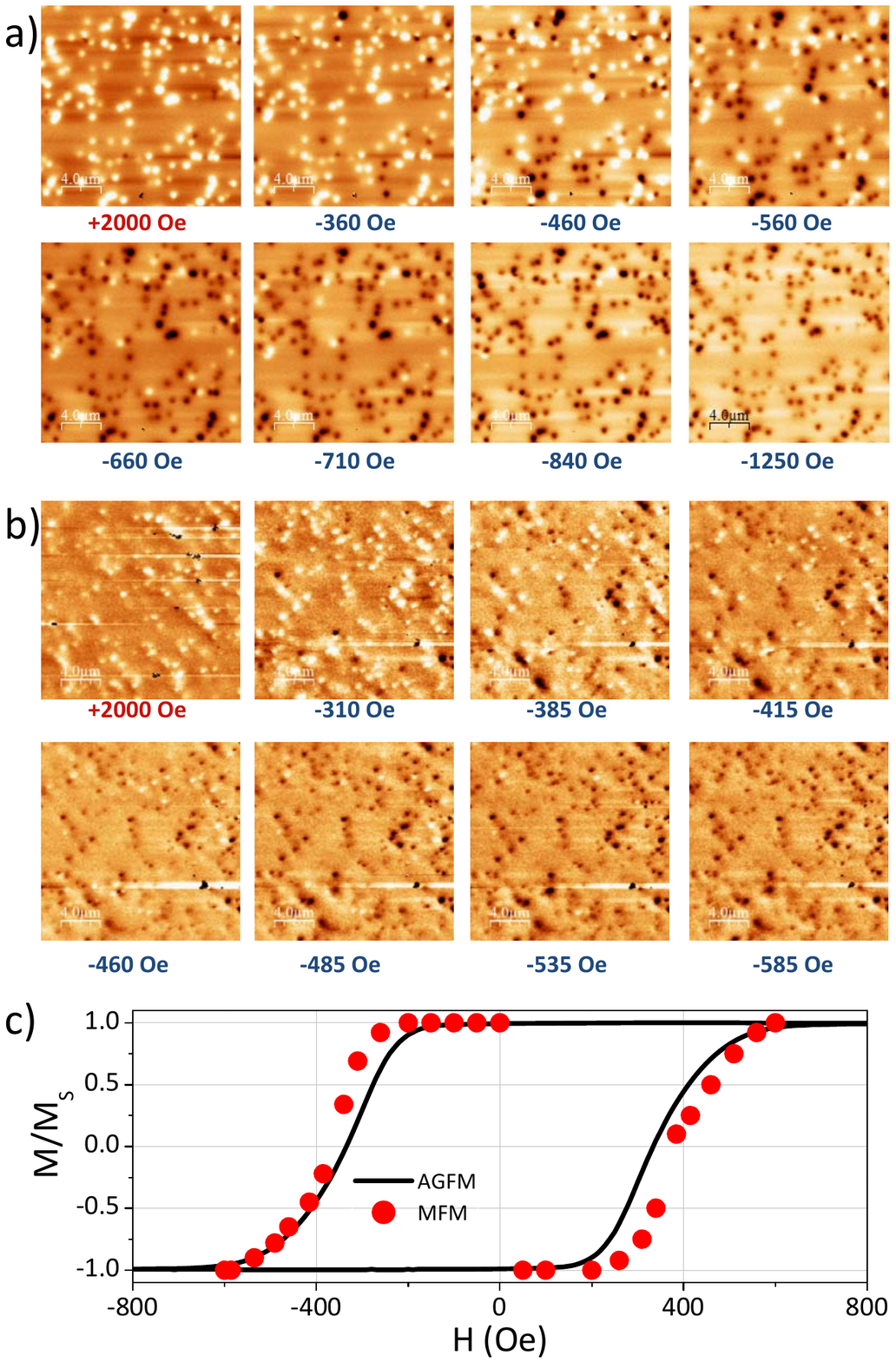}
\caption{MFM images of the Co$_{55}$Fe$_{45}$ nanowire array ($D=70$ nm)
at zero field after saturation in a positive 2 kOe field and then
in a variety of non-saturated remnant states obtained by changing
each time the magnetic state of the nanowire array from saturation
to a given negative field and then to remanence by reducing the field
to zero. All the images were scanned in the same area; (b) same as
in (a) for the Ni$_{80}$Fe$_{20}$ nanowire array ($D=7$0 nm) (c)
Hysteresis loop of the Ni$_{80}$Fe$_{20}$ nanowire array ($D=70$
nm) measured by AGFM (solid line) and MFM (red dots) with the applied
field parallel to the wire axis.}
\end{figure}

These images first demonstrate the bistable behaviour of the nanowires.
The MFM investigations show the magnetization at the top of the NW
and the presence of only two magnetization states indicates that they
are bistable and single domain. Indeed, the bright and the dark contrasts
of the NW are the result of their magnetization antiparallel or parallel
to the magnetization of the probe, respectively, as explained above.
Furthermore, the coercivity of the probe used is 400 Oe whereas the
moment is $3\times10^{-13}$ emu so its stray field is not influencing
the magnetization reversal of the arrays without applying external
magnetic field. From the MFM images, it was confirmed that only two
remnant states are stable inside each NW corresponding to single domain
NW with \textquotedblleft{}up\textquotedblright{} (along $+Oz$) and
\textquotedblleft{}down\textquotedblright{} (along $-Oz$) magnetisation
directions. Consequently, the local hysteresis curve was calculated
by counting the number of NW with \textquotedblleft{}up states\textquotedblright{}
and \textquotedblleft{}down states\textquotedblright{}. Thus, the
normalized magnetization can be written as {[}12{]}: 
\[
\frac{M^{MFM}(H)}{M_{S}^{MFM}}=\frac{n_{up}-n_{down}}{n_{up}+n_{down}}
\]

Where $n_{up}$ and $n_{down}$correspond to the number of NW in up
and down states in the image; $M_{S}^{MFM}$is the saturation magnetization
and $M^{MFM}(H)$ is the magnetization at field $H$; note that the
coercive field $H_{C}$ is the field at which $n_{up}=n_{down}$.
Figure 2c presents a comparison between the MFM-magnetization curve
(red dots) and the magnetization curve from AGFM (continuous line)
measured for the array of Ni$_{80}$Fe$_{20}$ NW ($D=$ 70 nm). 

Actually, the measurements performed by a magnetometer are sensitive
to the entire volume of the NW whereas the MFM measurements are not
sensitive to it since it is a surface imaging technique. From the
good agreement between both kinds of measurements, it can be assumed
that the scanned area ($20\times20$ $\lyxmathsym{\textmu}$m2) is
large enough to ensure that the effect of NW volume distribution is
taken into account. In addition, MFM allows mapping the local magnetization
curves for the NW arrays with very low packing density where the volume
magnetic moment is too weak to be measured by bulk magnetometry.

From the images presented in Figures 2a-b, we observe that the magnetization
reversal of the NW is nearly random. This supports the assumption
of weak dipolar interactions. Indeed, in AAO arrays of NW {[}9, 19{]}
exhibiting strong dipolar interactions, the switching of one NW is
strongly influenced by the magnetization state of its neighbours.
In addition, the demagnetized state of AAO NW arrays presents a labyrinth
domain pattern, as expected for a hexagonal array to maintain neighbouring
bits with antiparallel magnetization and to minimize the inter NW
dipolar field energy {[}9{]}. The demagnetized state of the present
arrays corresponds roughly $50\%$ of black (resp. white) spots, randomly
distributed (not shown here) which is consistent with an array presenting
negligible dipolar interactions.

\begin{figure}
\includegraphics[bb=30bp 105bp 370bp 590bp,clip,width=8.5cm]{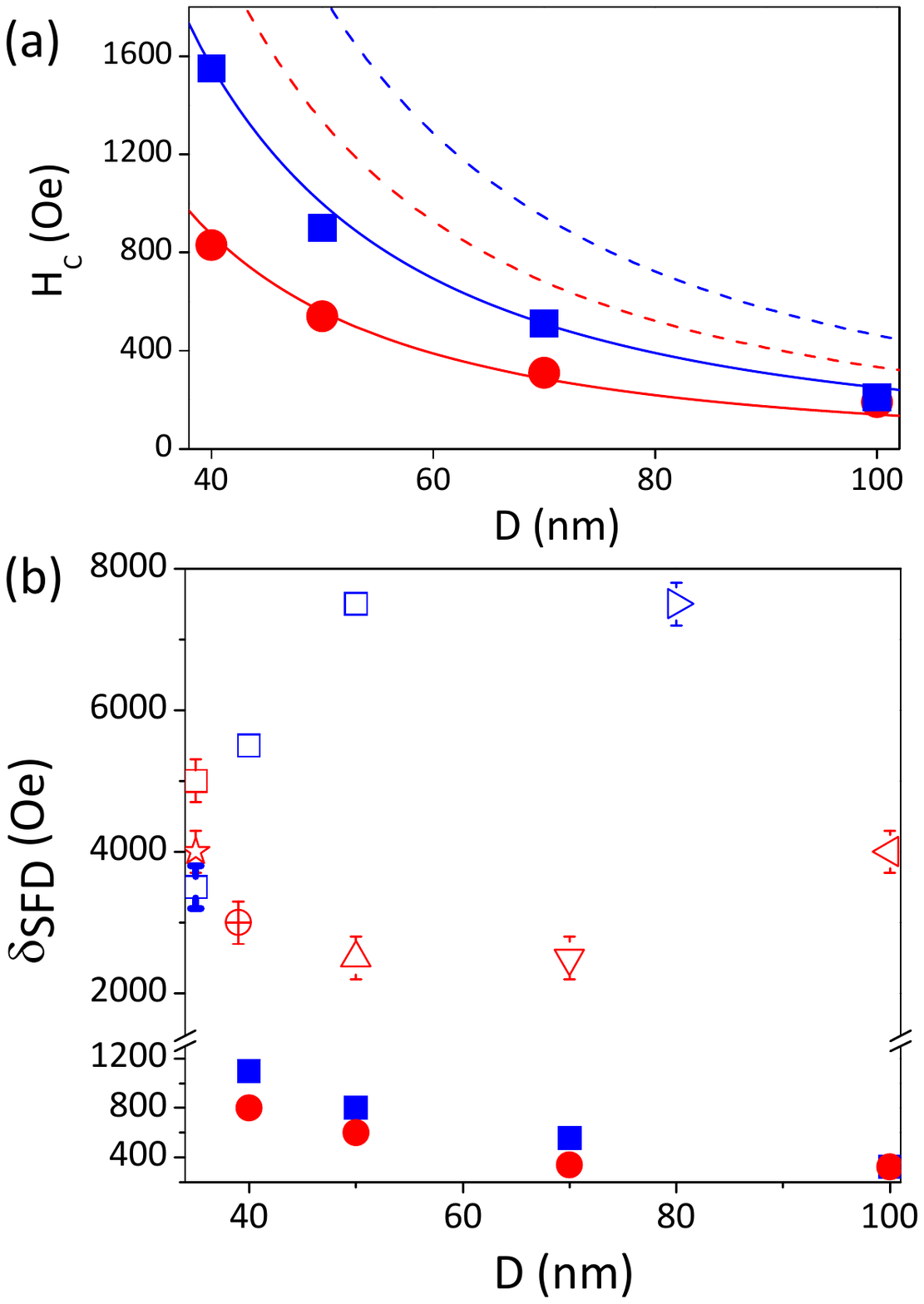}
\caption{a) Variation of the average coercive field as a function of diameter
for the Ni$_{80}$Fe$_{20}$ (circles) and Co$_{55}$Fe$_{45}$ (squares)
NW arrays; the dashed lines showing a $D^{-2}$ variation are calculated
by using the analytical formula of $H_{C}$ in the curling model {[}27{]};
the continuous lines are the fits obtained by using a $D^{-2}$ law.
b) Variation of the switching field distribution width of Ni$_{80}$Fe$_{20}$
and Co$_{55}$Fe$_{45}$ NW arrays (filled circles and squares, respectively)
as a function of $D$ and comparison with previous results obtained
on NW arrays embedded in AAO templates; blue open symbols represent
Co$_{55}$Fe$_{45}$ NWs whereas the red open symbols represent Ni$_{80}$Fe$_{20}$
NW: {[}20, 23-27{]} and {[}3, 16{]}, respectively. }
\end{figure}

The coercive field measured by MFM is reported in Figure 3a: it increases
with decreasing $D$ for the two sets of NW arrays. This variation
can be well fitted using a $D^{-2}$ law (solid lines in Figure 3a).
Magnetization reversal analytical models of an infinite cylinder have
been used to qualitatively explain this dependence {[}28{]}. It is
found that the dependence of $H_{C}$ on the diameter of the nanowires
in the curling model, as given by $H_{C}=6.78A\left(M_{S}R^{2}\right)^{-1}$
, where $A$ is the exchange stiffness and $R$ the radius of the
NW is in qualitative agreement with the experimental observations.
However, the values found with this model are approximately two times
larger than the experimental ones. Such difference has already been
found in previous studies where the $D$-dependence of the coercive
field of non-interacting Ni NWs embedded in PC membranes has been
experimentally determined and compared to micromagnetic simulations
and analytical models {[}29{]}. Shape deviations from the ideal cylinder
{[}30{]} like structural defects at the extremity of the NW can play
the role of nucleation sites for the magnetization reversal and may
be the origin of these differences. However, from the $D$-dependence
of the coercive field, it can be concluded that in these arrays of
NW, the magnetization reversal follows the curling mode.

The SFD is linked with the number of switched NW after each increment
of the applied magnetic field. The SFD width ($\delta_{SFD}$) which
corresponds to the field range between the magnetization reversal
of the first NW and that of the last one is also found to be $D$-dependent
(see Figure 3b). The $\delta_{SFD}$ obtained from densely packed
NW arrays are reported in Figure 3b for comparison. These values are
at least 3 times larger than the ones obtained in this study {[}3,9,11,
14, 16,20, 22- 27{]}. In the dense arrays, the dipolar interactions
are strong and lead to a broadening of $\delta_{SFD}$ {[}9{]}, whereas,
in the present study $\delta_{SFD}$ is essentially due to intrinsic
effects and no complex calculation is required to discriminate between
the intrinsic and dipolar contributions to the switching field distribution.
Wang et \textit{al}. {[}11{]} reported the intrinsic SFD in an array
of Co NW ($D=$ 30 nm) embedded in AAO membranes by evaluating and
subtracting the total dipolar field present in this array. Even after
subtracting the effect of this dipolar field, an intrinsic SFD width
of about 4 kOe was estimated (i. e. about one order of magnitude larger
than in the present study) which confirms the complexity of such an
analysis. The deviation of $\delta_{SFD}$ from an ideal Dirac function
may be due to parameters such as the distributions of diameter and
length, the non-uniformity of the shape, inhomogeneities in the microstructure
and chemical composition of the NW. 

To explain the variation of the $\delta_{SFD}$ with $D$ (see Figure
3b), the size distribution of the nanopore diameter which can be fitted
by a Gaussian law with a standard deviation of $\sigma_{D}\simeq\pm5$
nm has been taken into account. Assuming that $\delta_{SFD}$ can
be written as $\delta_{SFD}=\delta_{SFD}^{D}+\delta_{SFD}^{0}$ where,
$\delta_{SFD}^{D}$ is the contribution of the diameter distribution
and $\delta_{SFD}^{0}$ refers to the contribution of other factors
such as non-uniform NW lengths and tips and inhomogeneities in the
microstructure and chemical composition. The two major contributions
$\delta_{SFD}^{D}$ and $\delta_{SFD}^{0}$ can be determined separately.
For calculating $\delta_{SFD}^{D}$ , the product of diameter distribution
and the $D^{-2}$ fits obtained from the experimental evolutions of
$H_{C}$ with $D$ (Figure 3a), from MFM measurements, have been used.

\begin{figure}
\includegraphics[bb=30bp 160bp 355bp 590bp,clip,width=8.5cm]{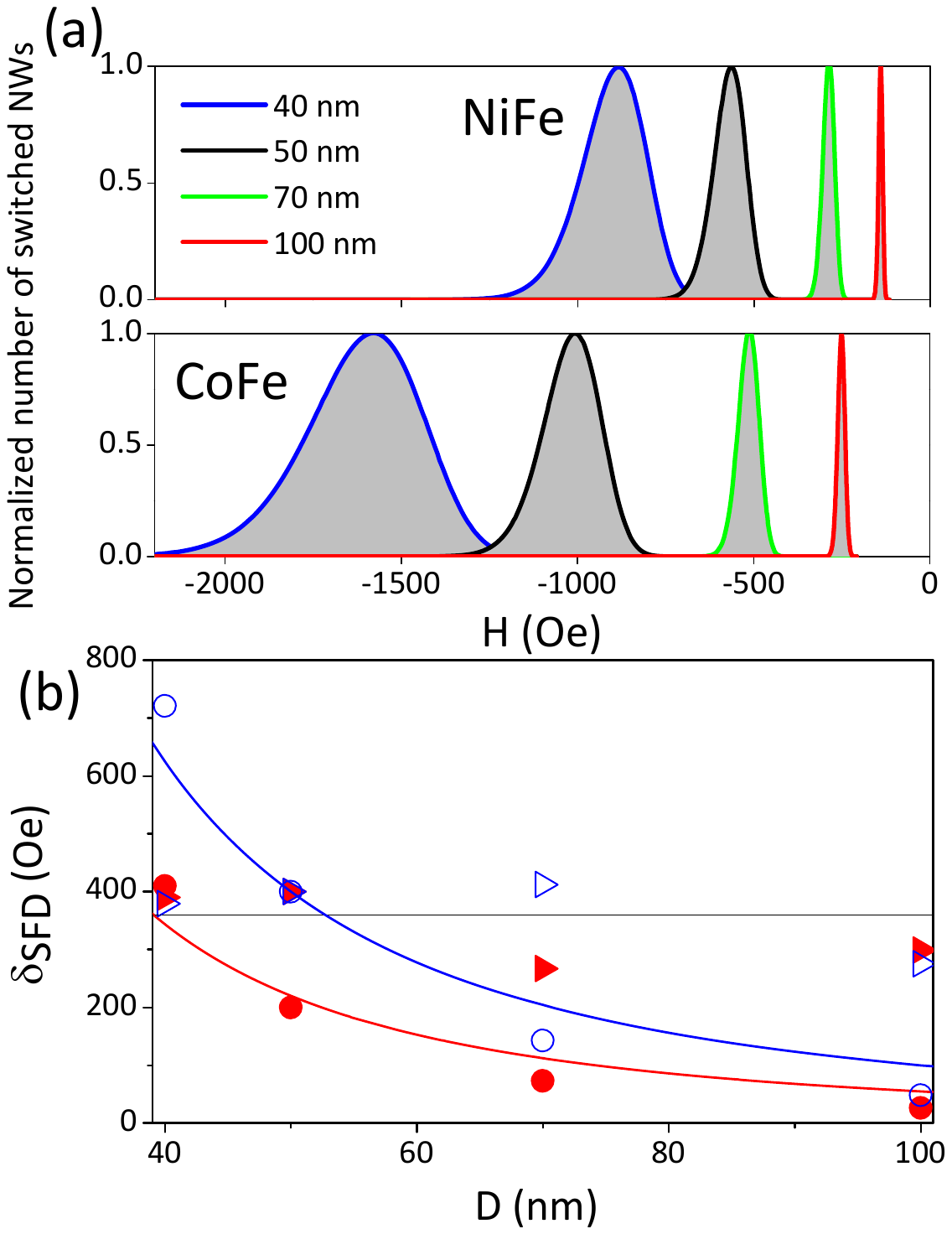}
\caption{a) Switching field distribution of Ni$_{80}$Fe$_{20}$ and Co$_{55}$Fe$_{45}$
nanowire arrays considering the Gaussian distribution of pore sizes
for diameters in the range 40-100 nm. b) Variation of the switching
field distribution width as a function of diameter (circles) and its
variation as obtained by separating the pore size distribution from
the intrinsic contribution (triangles); the continuous lines are guide
for the eyes (open symbols and filled symbols for Co$_{55}$Fe$_{45}$
and for Ni$_{80}$Fe$_{20}$, respectively).}
\end{figure}

Figure 4a presents the calculated in this way while Figure 4b shows
that increases with decreasing $D$ and is well fitted using a $D^{-2}$
law. Note that the appears to be almost independent of $D$ for the
two NW arrays (triangles in Figure 4b) and is around the value of
350 Oe, which is quite low for our NW arrays in comparison with AAO
NW arrays. This means that the significant contribution to the total
SFD of AAO NW arrays and also of bit patterned media {[}30{]} originates
from non-intrinsic dipolar interaction effects. The measured in the
present study is essentially intrinsic. One may think that the broader
$\delta_{SFD}$ for smaller diameters could be the consequence of
an increased packing density and hence stronger dipolar interactions.
Nevertheless, in the arrays under study, $P$ slightly enhances with
increasing $D$ whereas $\delta_{SFD}$ is decreasing showing that
influence of the $\sigma_{D}$ on $\delta_{SFD}$ was more prominent
for small $D$. If dipolar interactions were present, this would have
led to the broadening of $\delta_{SFD}$ when $D$ increases which
is not the case in present measurements.

\section{Conclusion}

The magnetization reversal process in arrays of Ni$_{80}$Fe$_{20}$
and Co$_{55}$Fe$_{45}$ ferromagnetic NW embedded into low porosity
nanoporous polycarbonate membranes has been studied systematically
by \textit{in situ} magnetic force microscopy. The fabricated NWs
have high aspect ratio and negligible magnetocrystalline anisotropy,
so the magnetization has been found to be parallel to the NW axis
displaying bistable magnetic behaviour. By analysing the numbers of
NWs with magnetization up and down, local magnetization curves were
obtained which agree well with bulk magnetization measurements from
AGFM. The images taken under continuous magnetic field and in the
remnant state were compared and found to be identical demonstrating
that for these low density arrays the remnant state is identical to
the in-field state due to the weak dipolar interaction. The influence
of the diameter on the magnetization reversal and the SFD is shown.
Regardless of the NW diameter, narrow SFD widths ($\delta_{SFD}$
) are observed compared to the ones generally obtained for densely
packed NW arrays in AAO membranes (a SFD width of 300 Oe was obtained
for the 100 nm samples). Moreover, an increase of $\delta_{SFD}$
with decreasing diameter is observed and is qualitatively explained
by considering the size distribution of the template nanopores. 
\begin{acknowledgments}
The authors thank E. Ferain for providing the PC templates and Pascal
Van Velthem for technical assistance. J. De La Torre and Armando Encinas
thanks CONACYT for financial support through grants No. 166089, 177896,
105568 and 162651. M.R. Tabasum and B. Nysten acknowledge financial
support of the Belgian government in the framework of the \textquotedblleft{}Interuniversity
Attraction Poles\textquotedblright{} program (IUAP7/6-FS2). M.R Tabasum
is Assistant Professor on leave from department of industrial and
manufacturing engineering RCET-UET Lahore.\end{acknowledgments}

\end{document}